\documentclass[sn-vancouver,default,iicol]{sn-jnl}

\usepackage{natbib}

\jyear{2022}%

\theoremstyle{thmstyleone}%
\usepackage{mathtools}
\usepackage{amssymb}
\usepackage{multirow}
\usepackage{amsmath}
\usepackage{braket}
\usepackage{mathrsfs}
\theoremstyle{thmstyletwo}%

\theoremstyle{thmstylethree}%

\begin{document}

\title[]{Accurate Temperature Diagnostics for Matter under Extreme Conditions}


\author*[1,2]{\fnm{Tobias} \sur{Dornheim}}\email{t.dornheim@hzdr.de}

\author[1,2,3]{\fnm{Maximilian} \sur{B\"ohme}}

\author[4,2]{\fnm{Dominik} \sur{Kraus}}

\author[5]{\fnm{Tilo} \sur{D\"oppner}}

\author[6]{\fnm{Thomas~R.} \sur{Preston}}

\author[1,2]{\fnm{Zhandos~A.} \sur{Moldabekov}}

\author[2]{\fnm{Jan} \sur{Vorberger}}

\affil[1]{\orgname{Center for Advanced Systems Understanding (CASUS)}, \orgaddress{\city{G\"orlitz}, \postcode{D-02826}, \country{Germany}}}

\affil[2]{\orgname{Helmholtz-Zentrum Dresden-Rossendorf (HZDR)}, \orgaddress{\city{Dresden}, \postcode{D-01328},  \country{Germany}}}

\affil[3]{\orgname{Technische  Universit\"at  Dresden}, \orgaddress{\city{Dresden}, \postcode{D-01062}, \country{Germany}}}

\affil[4]{\orgname{Institut f\"ur Physik, Universit\"at Rostock}, \orgaddress{\city{Rostock}, \postcode{D-18057}, \country{Germany}}}

\affil[5]{\orgname{Lawrence Livermore National Laboratory}, \orgaddress{\city{Livermore}, \postcode{California 94550}, \country{USA}}}

\affil[6]{\orgname{European XFEL}, \orgaddress{\city{Schenefeld}, \postcode{D-22869}, \country{Germany}}}


\abstract{ 
The experimental investigation of matter under extreme densities and temperatures~\cite{Hatfield_Nature_2021} as they occur for example in astrophysical objects~\cite{Kritcher_Nature_2020,Liu2019,Bailey2015,Brygoo2021,Kraus_Science_2022} and nuclear fusion applications~\cite{Zylstra2022,Betti2016,Hurricane_Nature_2014} constitutes one of the most active frontiers at the interface of material science, plasma physics, and engineering~\cite{Fletcher2015,Knudson_Science_2015,Kraus2016,Kraus2017,Dias_Silvera_Science_2017,Celliers_Science_2018,Lazicki2021}. The central obstacle is given by the rigorous interpretation of the experimental results~\cite{siegfried_review}, as even the diagnosis of basic parameters like the temperature $T$ is rendered highly difficult by the extreme conditions. In this work, we present a simple, approximation-free method to extract the temperature of arbitrarily complex materials from scattering experiments, without the need for any simulations or an explicit deconvolution. 
This new paradigm can be readily implemented at modern facilities and corresponding experiments will have a profound impact on our understanding of warm dense matter and beyond, and open up a gamut of appealing possibilities in the context of thermonuclear fusion, laboratory astrophysics, and related disciplines.
}




\maketitle

The study of matter at extreme conditions (temperatures of $T\sim10^4-10^8$K and pressures of $P\sim1-10^4$Mbar) constitutes one of the most fundamental challenges of our time~\cite{Hatfield_Nature_2021}. Such \emph{warm dense matter} (WDM)~\cite{wdm_book,new_POP} is ubiquitous throughout our Universe~\cite{fortov_review} and naturally occurs in a number of astrophysics objects such as giant planet interiors~\cite{Liu2019,Brygoo2021,Kraus_Science_2022}, white and brown dwarfs~\cite{Kritcher_Nature_2020,becker}, and the outer layer of neutron stars~\cite{neutron_star_envelopes}. On Earth, WDM can nowadays be realized experimentally at large research facilities using different techniques~\cite{falk_wdm}, and particularly advantageous photon properties are offered by x-ray free electron lasers such as LCLS, European XFEL, or SACLA~\cite{Yabashi2017,Tschentscher_2017,LCLS_2016,SACLA_2011}. This opens up enticing new possibilities for laboratory astrophysics~\cite{takabe_kuramitsu_2021}, the discovery of novel materials~\cite{Lazicki2021,Kraus2017}, and hot-electron chemistry~\cite{Brongersma2015}.
A particularly important application is given by inertial confinement fusion (ICF)~\cite{Zylstra2022,Betti2016,Hurricane_Nature_2014}, which promises a potential abundance of clean energy in the future. Here the fuel capsule is traversing the WDM regime on its path towards ignition~\cite{hu_ICF}. 
Consequently, a number of experimental breakthroughs~\cite{Fletcher2015,Knudson_Science_2015,Kraus2016,Kraus2017,Dias_Silvera_Science_2017,Celliers_Science_2018,Lazicki2021} have been reported over the last years.

Unfortunately, the rigorous diagnosis of such experiments is rendered highly demanding by the extreme conditions. Indeed, even basic properties such as the temperature, that can be considered as well known in many other experiments, cannot be directly measured at WDM conditions and have to be inferred from other observations~\cite{kraus_xrts}.
In this regard, the X-ray Thomson scattering (XRTS) technique~\cite{siegfried_review} has emerged as a promising method of diagnosis. Yet, the actual inference of the temperature from the experimentally measured XRTS signal is substantially hampered by three major obstacles. Firstly, the theoretical modelling of the dynamic structure factor $S(\mathbf{q},E)$ [with $\mathbf{q}$ being the wave vector and $E$ the energy transfer] of a real WDM system constitutes a most formidable challenge~\cite{wdm_book,new_POP,dornheim_dynamic}. In practice, one usually has to rely on uncontrolled approximations such as the Chihara decomposition~\cite{kraus_xrts,Chihara_1987}, or time-dependent density functional theory~\cite{Ramakrishna_PRB_2021}. In addition to their unclear accuracy, state-of-the-art simulations are computationally highly demanding, which makes them impractical for the required parameter optimization, and prevents their application to complex materials. Secondly, the experimentally measured XRTS signal is given by the convolution of $S(\mathbf{q},E)$ with the instrument function $R(E)$: $I(\mathbf{q},E) = S(\mathbf{q},E)\circledast R(E)$. Therefore, important features may be smeared out, and the direct usage of the \emph{detailed balance} relation on the scattered signal~\cite{DOPPNER2009182} $S(\mathbf{q},-E)=S(\mathbf{q},E)e^{-\beta E}$ (with the inverse temperature $\beta=1/k_\textnormal{B}T$) is not always possible. Thirdly, the experimental signal is always afflicted with statistical noise. This may further camouflage physical features, and usually prevents a deconvolution.

\begin{figure*}\centering
\includegraphics[width=0.39\textwidth]{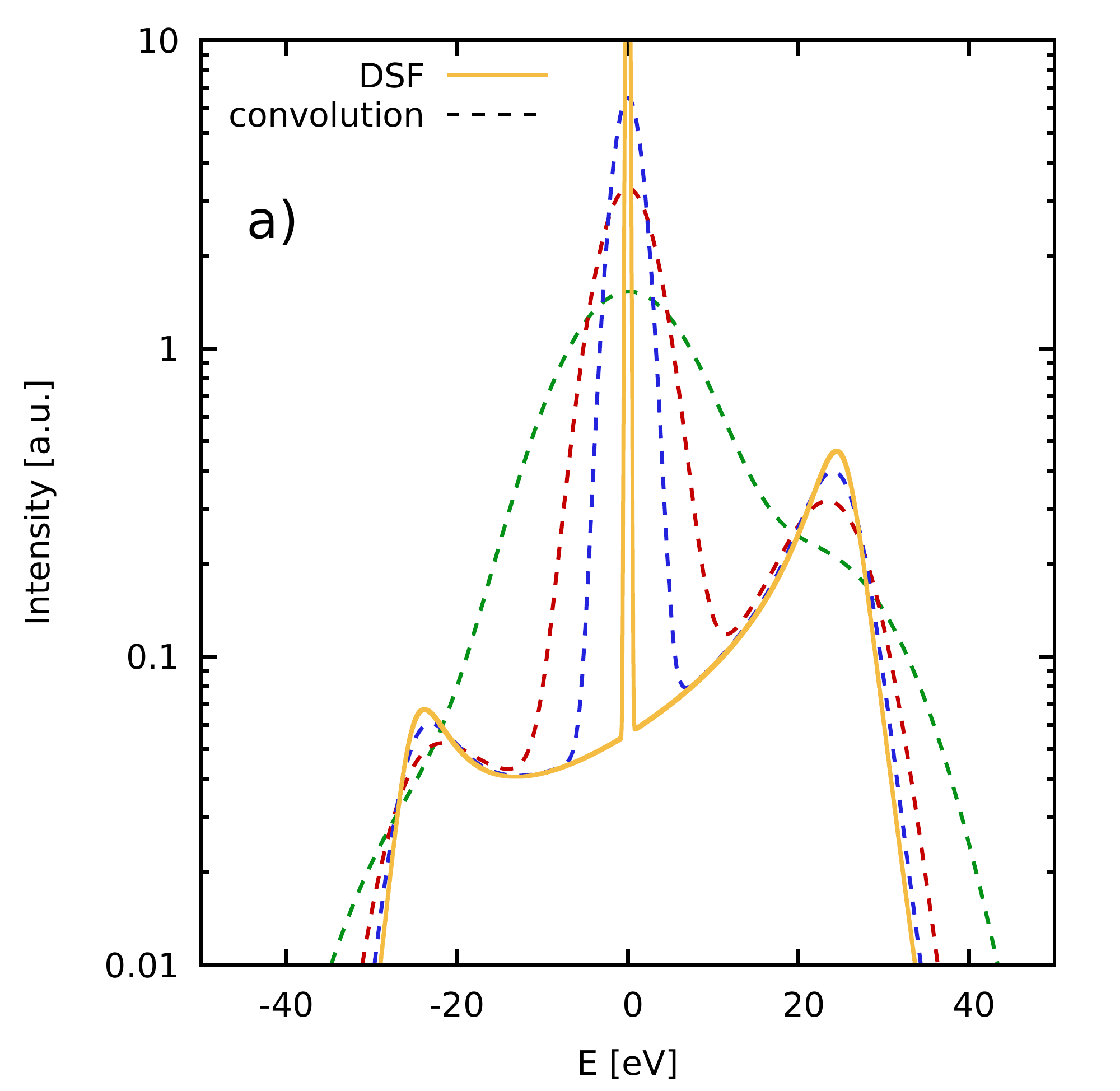}
\includegraphics[width=0.39\textwidth]{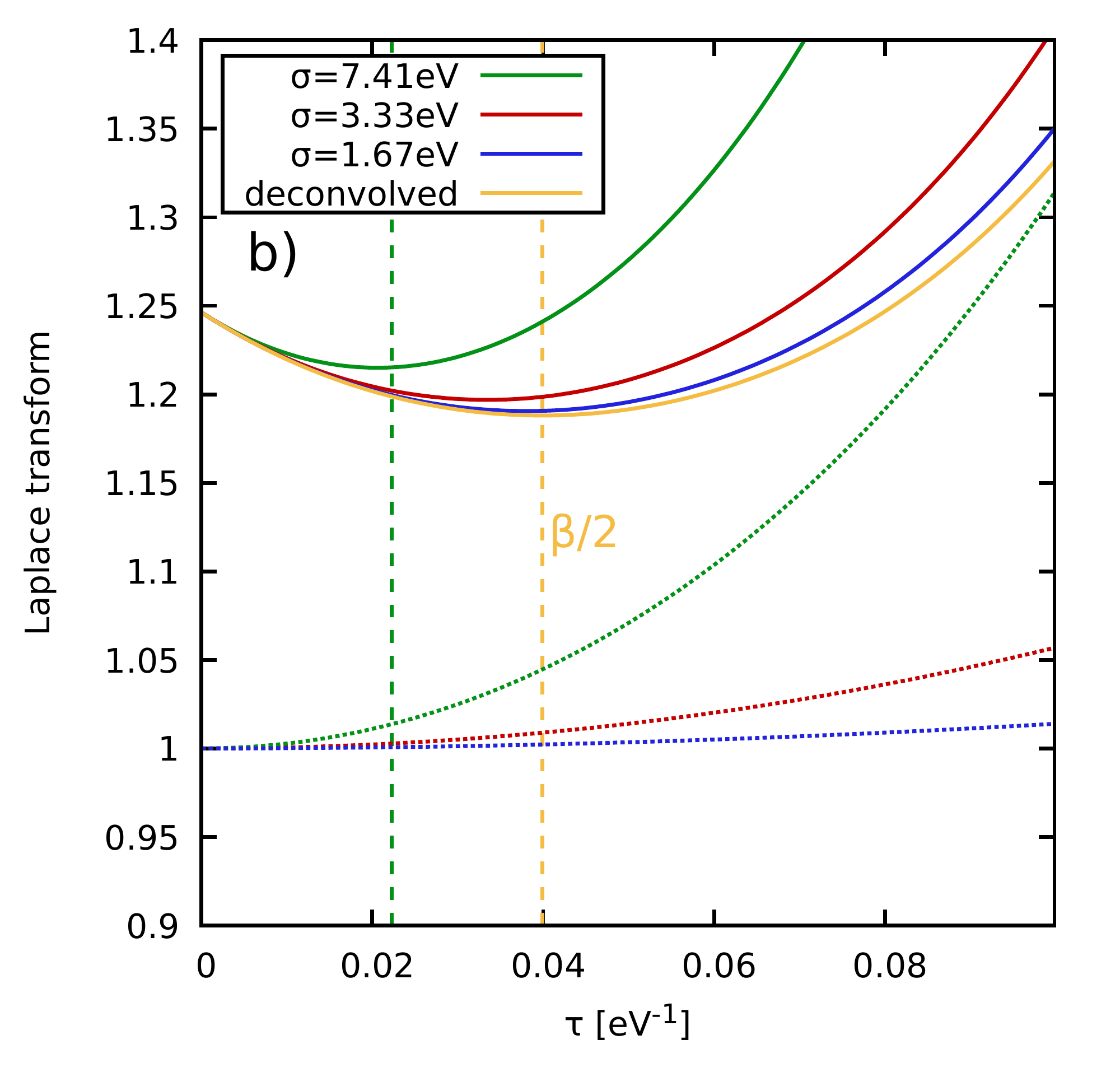}
\caption{\label{fig:synthetic} Demonstration of the temperature diagnostics from an XRTS signal. Panel a) shows
synthetic results for the dynamic structure factor (solid yellow) of a uniform electron gas~\cite{dornheim_dynamic} at the metallic density of $r_s=\overline{r}/a_\textnormal{B}=2$ (with $\overline{r}$ being the average electronic separation) and the electronic Fermi temperature $T=12.53\,$eV, for half the Fermi wave number~\cite{quantum_theory} $q=0.5q_\textnormal{F}=0.91$\AA$^{-1}$. The dashed curves have been convolved with a Gaussian instrument function with different widths $\sigma$. Panel b) shows the corresponding evaluation of the two-sided Laplace transform of the respective XRTS signals (solid), and instrument functions (dotted). Dividing the former by the latter gives the exact curve of $F(\mathbf{q},\tau)$ [yellow] corresponding to the deconvolved dynamic structure factor $S(\mathbf{q},E)$.
}
\end{figure*}

In this work, we present a complete and straightforward solution to all three obstacles. Specifically, we propose to analyse the two-sided Laplace transform [cf.~Eq.~(\ref{eq:ITCF})] of the measured XRTS signal, $\mathcal{L}\left[I(\mathbf{q},E)\right]$, which gives us direct and unbiased access to the temperature of the probed system. Our new approach is \emph{model-free}, computationally cheap and works for arbitrarily complex materials. This makes it very practical for the on-the-fly interpretation of experiments, and opens up new avenues to a plethora of applications such as the characterization of the ablator of an ICF fuel capsule, or the study of material mixtures at the conditions encountered within planets~\cite{Georg2007}.  
Moreover, the effect of the instrument function can be fully taken into account without the need for a numerically unstable explicit deconvolution. 
Finally, the definition of $\mathcal{L}\left[\dots\right]$ as an integration over the full relevant energy-range makes it very robust to noise in the experimentally measured signal.

To highlight the flexibility and practical value of our new methodology, we apply it to three representative XRTS experiments: i) the pioneering observation of plasmons in warm dense beryllium by Glenzer~\textit{et al.}~\cite{Glenzer_PRL_2007}; ii) the study of isochorically heated aluminium by Sperling~\textit{et al.}~\cite{Sperling_PRL_2015}, which has resulted in an ongoing controversy~\cite{Dornheim_PRL_2020_ESA,Mo_PRL_2018} regarding the nominal temperature of $T=6\,$eV; and iii) a recent  XRTS experiment with warm dense graphite by Kraus \textit{et al.}~\cite{kraus_xrts}, where standard interpretation models have resulted in uncertainties of $50\%$ with respect to $T$.
Our method works exceptionally well in all three cases and, in this way substantially reduces previous uncertainties.

\textbf{Idea.} Let us consider the two-sided Laplace transform of the dynamic structure factor:
\begin{eqnarray}\label{eq:ITCF}
\mathcal{L}\left[S(\mathbf{q},E)\right] = \int_{-\infty}^\infty \textnormal{d}E\ e^{-\tau E} S(\mathbf{q},E)\ .
\end{eqnarray}
In fact, Eq.~(\ref{eq:ITCF}) corresponds to the intermediate scattering function, $F(\mathbf{q},\tau)\equiv \mathcal{L}\left[S(\mathbf{q},E)\right]$, evaluated at imaginary times $-i\hbar\tau\in -i\hbar[0,\beta]$, which naturally emerges in Feynman's powerful path integral representation of statistical mechanics~\cite{Dornheim_JCP_ITCF_2021,kleinert2009path}.  $F(\mathbf{q},\tau) $ is symmetric around $\tau=\beta/2$ [cf.~Fig.~\ref{fig:synthetic} b)], see the Methods Section. This directly implies that knowledge of $S(\mathbf{q},E)$ gives straightforward access to the actual temperature of the system by solving the simple one-dimensional integral in Eq.~(\ref{eq:ITCF}), and subsequently locating the minimum in $F(\mathbf{q},\tau)$ at $\beta/2$.

An additional obstacle is given by the fact that the XRTS technique does not allow for measurements of $S(\mathbf{q},E)$, but its convolution with the instrument function $R(E)$. While the latter is typically known with high precision, the deconvolution of the measured intensity is generally rendered unstable by the statistical noise. Our idea completely circumvents this obstacle by exploiting the convolution theorem of $\mathcal{L}\left[\dots\right]$:
\begin{eqnarray}\label{eq:convolution}
\mathcal{L}\left[S(\mathbf{q},E)\right] = \frac{\mathcal{L}\left[S(\mathbf{q},E)\circledast R(E)\right]}{\mathcal{L}\left[R(E)\right]}\ .
\end{eqnarray}
In practice, we thus compute the two-sided Laplace transform of the experimentally measured intensity, which due to its definition as an integral, is very robust with respect to noise. The impact of the instrument function is then completely removed by the denominator of Eq.~(\ref{eq:convolution}), which also can be computed in a straightforward way. 
As a result, we get the unbiased temperature of any given system from the experimentally measured XRTS signal without the need for theoretical or computational models, and without any bias from the broadening due to the instrument function.

\begin{figure*}\centering
\hspace*{-0.0cm}\includegraphics[width=0.3249\textwidth]{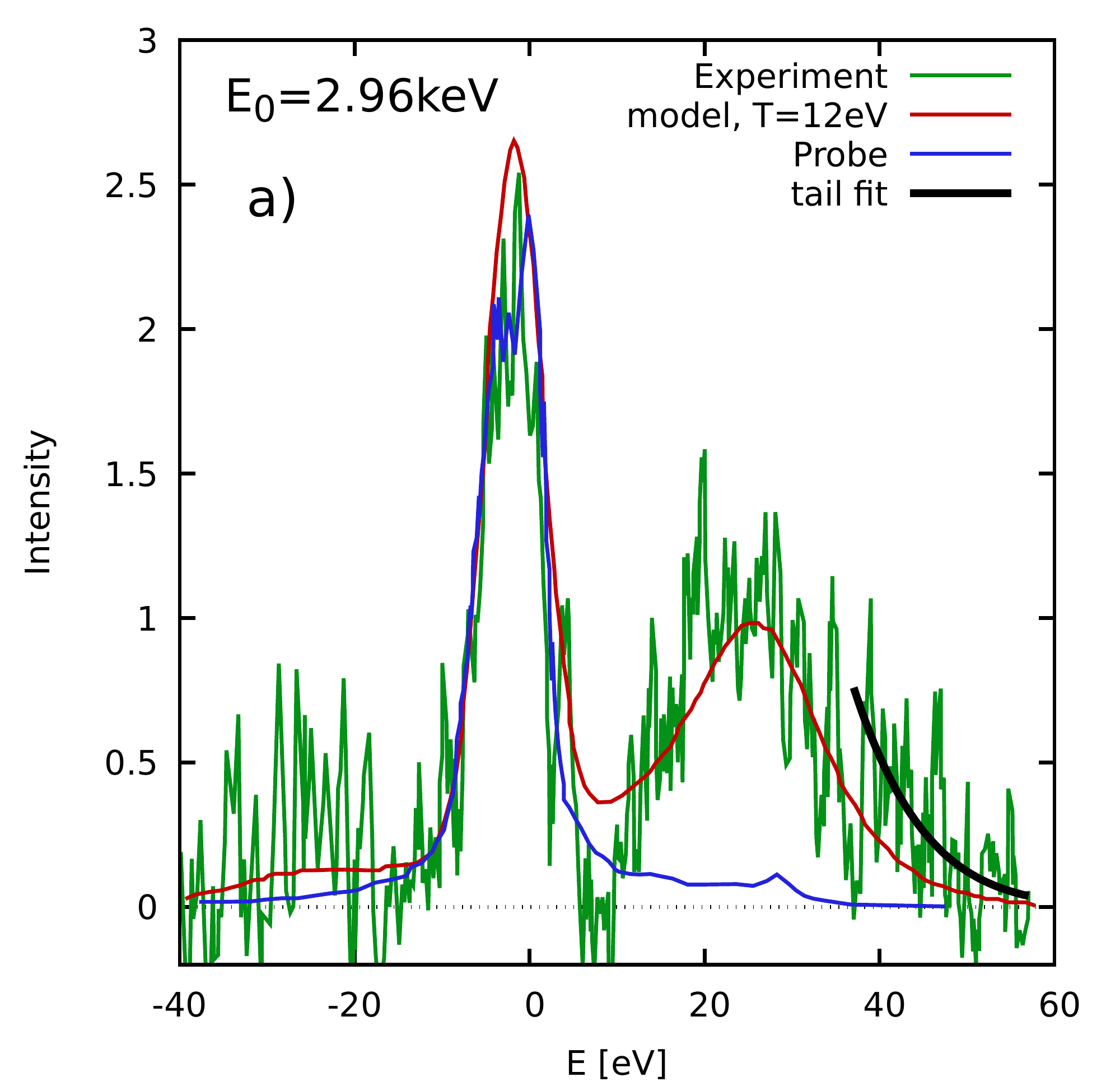}
\includegraphics[width=0.3249\textwidth]{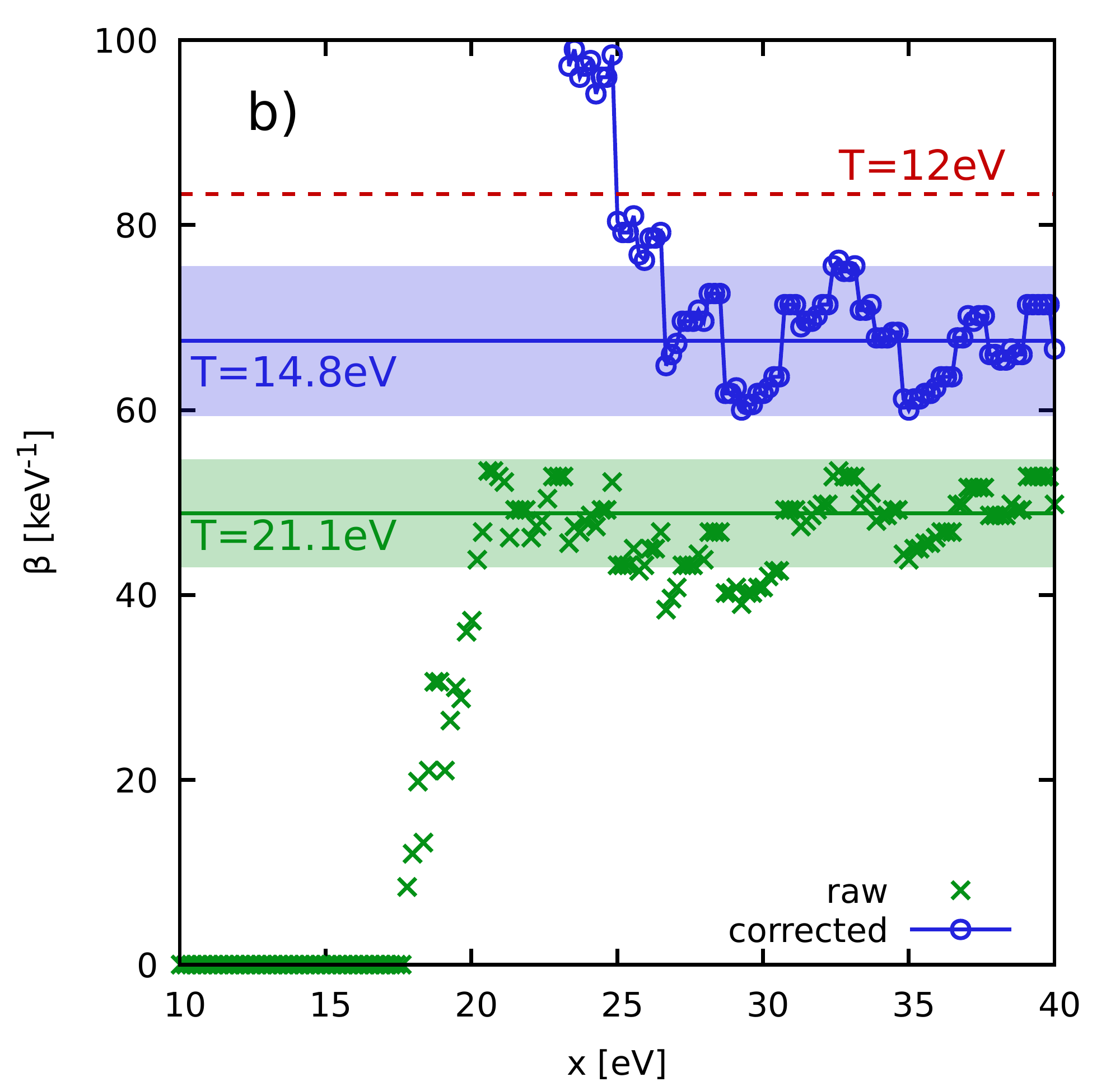}
\includegraphics[width=0.3249\textwidth]{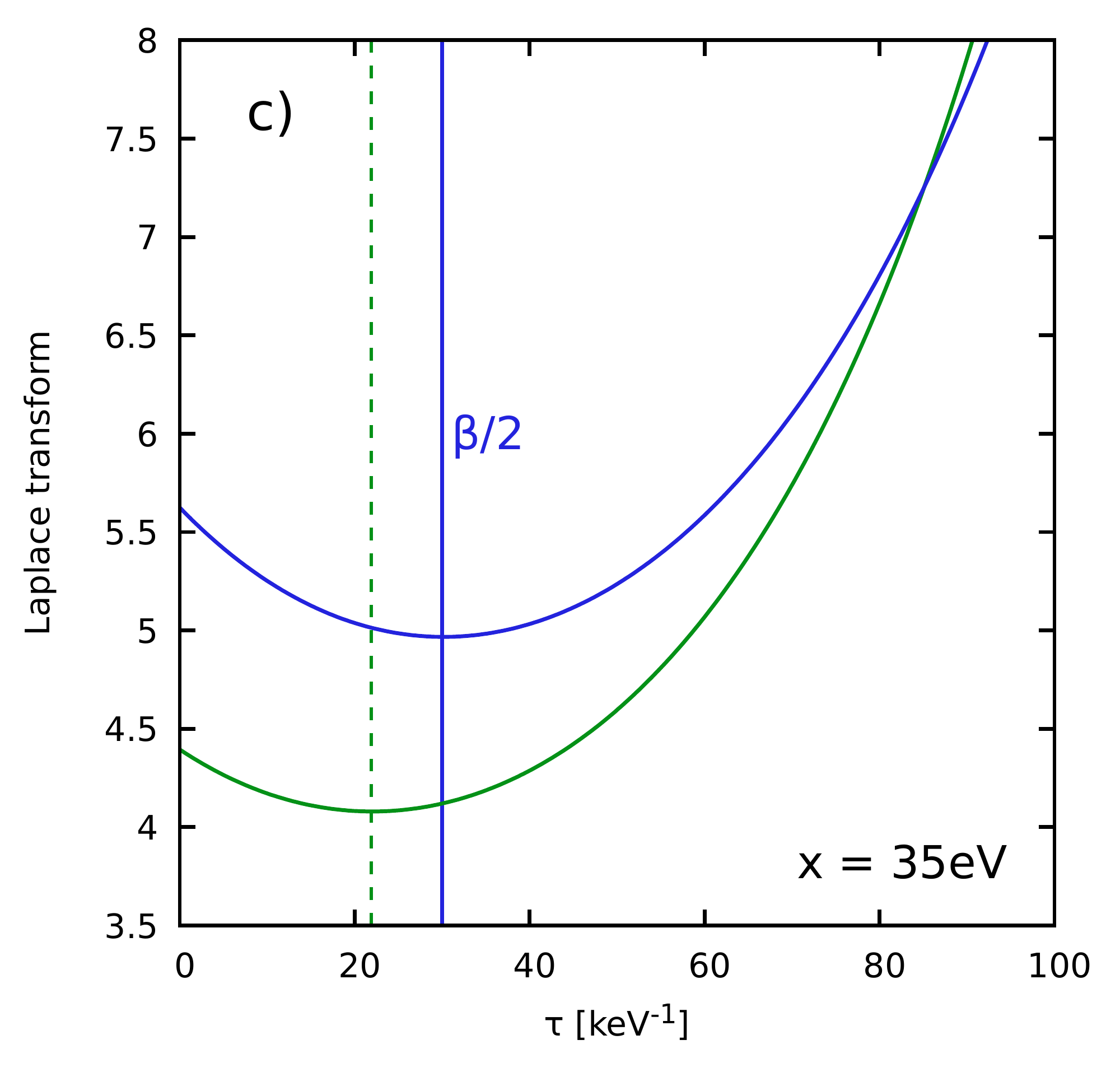}
\caption{\label{fig:Glenzer}
Temperature diagnosis of warm dense beryllium: a) XRTS measurement by Glenzer \emph{et al.}~\cite{Glenzer_PRL_2007} (green), theoretical Mermin model giving $T_\textnormal{model}=12\,$eV (red), instrument function $R(E)$ (blue), and averaged tail for $E\geq40\,$eV (black); b) convergence of our model-free temperature diagnosis with respect to the integration boundary $x$ of $\mathcal{L}\left[S(\mathbf{q},E)\right]$ with (blue) and without (green) correcting for the instrument function; c) corresponding results for the imaginary-time intermediate scattering function $F(\mathbf{q},\tau)$.
}
\end{figure*}

\textbf{Results.} To demonstrate our new methodology, we consider synthetic data in Fig.~\ref{fig:synthetic}. In panel a), we show the XRTS intensity based on a uniform electron gas model~\cite{review,dornheim_dynamic,quantum_theory} (with an additional sharp elastic peak around $E=0$) at a metallic density (Wigner-Seitz radius~\cite{quantum_theory} $r_s=\left(3/4\pi n_e\right)^{1/3}=2$, with $n_e$ being the electron density; this is close to both beryllium and aluminium) at the electronic Fermi temperature of $T=12.53\,$eV and half the Fermi wave number, i.e., $q=0.91$\AA$^{-1}$. 
In particular, the solid yellow curve shows $S(\mathbf{q},E)$, and the dashed curves have been obtained by convolving the latter with Gaussian instrument functions of different realistic widths $\sigma$. With increasing $\sigma$, the thus synthesised intensities become broader, and the plasmon peaks around $E=\pm25\,$eV are smeared out. It is important to note that the convolved curves do \emph{not} fulfill the aforementioned detailed balance relation between positive and negative energies, so that a direct extraction of the given temperature from such a dataset is not possible.

Let us next consider the corresponding evaluation of the different ingredients to Eq.~(\ref{eq:convolution}), which are shown in Fig.~\ref{fig:synthetic} b). The solid yellow line corresponds to the actual imaginary-time intermediate scattering function $F(\mathbf{q},\tau)=\mathcal{L}\left[S(\mathbf{q},E)\right]$, which has a minimum at $\tau=\beta/2$, see the vertical dashed yellow line. Evaluating the two-sided Laplace transform of the convolved curves give the solid blue, red, and green curves, which noticeably deviate from the exact $F(\mathbf{q},\tau)$. This is a direct consequence of the broadening due to the instrument function $R(E)$, leading to a violation of the detailed balance. Evidently, considering the minimum of the Laplace transform of the convolved signal leads to a substantial overestimation of the temperature (i.e., an underestimation of the inverse temperature $\beta$), as can be seen particularly well in the case of $\sigma=7.41\,$eV (vertical dashed green line). 

To remove the bias due to the instrument function, we have to compute the denominator $\mathcal{L}\left[R(E)\right]$ of Eq.~(\ref{eq:convolution}), which is shown by the respective dotted curves. Indeed, dividing the solid lines by the dotted lines in Fig.~\ref{fig:synthetic} b) recovers the true $F(\mathbf{q},\tau)$---and, therefore, the actual value of the temperature $T$---for all cases.
We stress that our methodology works over the entire range of wave vectors $\mathbf{q}$, including the collective and single-particle regimes. In particular, no explicit resolution of a distinct plasmon peak in the experimentally measured intensity is required.

\textbf{Beryllium experiment.} As a first practical application of our new diagnostic methodology, we re-examine the pioneering observation of plasmons in warm dense beryllium by Glenzer \emph{et al.}~\cite{Glenzer_PRL_2007} in Fig.~\ref{fig:Glenzer}. Panel a) shows the measured XRTS signal (green) together with the instrument function $R(E)$ (blue) and a theoretical Mermin model~\cite{Mermin_model} that has been used in Ref.~\cite{Glenzer_PRL_2007} to infer the nominal temperature of $T_\textnormal{model}=12\,$eV. Panel b) shows the temperature as it has been computed from our method both from the convolved signal (green) and by additionally taking into account the instrument function via Eq.~(\ref{eq:convolution}) (blue). It is important to note that, in actual experiments, one only has the intensity over a finite range of energies, $E\in[E_\textnormal{min},E_\textnormal{max}]$. We thus truncate the integration boundaries of $\mathcal{L}\left[S(\mathbf{q},E)\circledast R(E)\right]$ at $\pm x$, and the corresponding results clearly converge around $x\gtrsim 30\,$eV despite the substantial noise in the experimental data.

We extract a temperature of $T=14.8\pm1.5\,$eV from the experimental data, which is close to, though significantly different from the value of $T_\textnormal{model}=12\,$eV (dashed red line) that has been inferred from the theoretical model. At the same time, we note that the higher temperature from our model-free diagnosis is consistent with the hydrodynamic simulations employed in the original Ref.~\cite{Glenzer_PRL_2007}. Moreover, it very plausibly fits to the XRTS signal shown in Fig.~\ref{fig:Glenzer} a), as the red curve noticeably underestimates the averaged tail for $E\gtrsim 40\,$eV (solid black). 
For completeness, we note that not taking into account the instrument function would result in the spurious temperature of $T=21.1\,$eV, see the green line in Fig.~\ref{fig:Glenzer} b).
Finally, the corresponding results for Eq.~(\ref{eq:convolution}) with and without the correction due to $R(E)$ are shown in panel c), which nicely illustrates the robustness of the Laplace transform with respect to noisy input data.

\begin{figure*}\centering
\hspace*{-0.0cm}\includegraphics[width=0.39\textwidth]{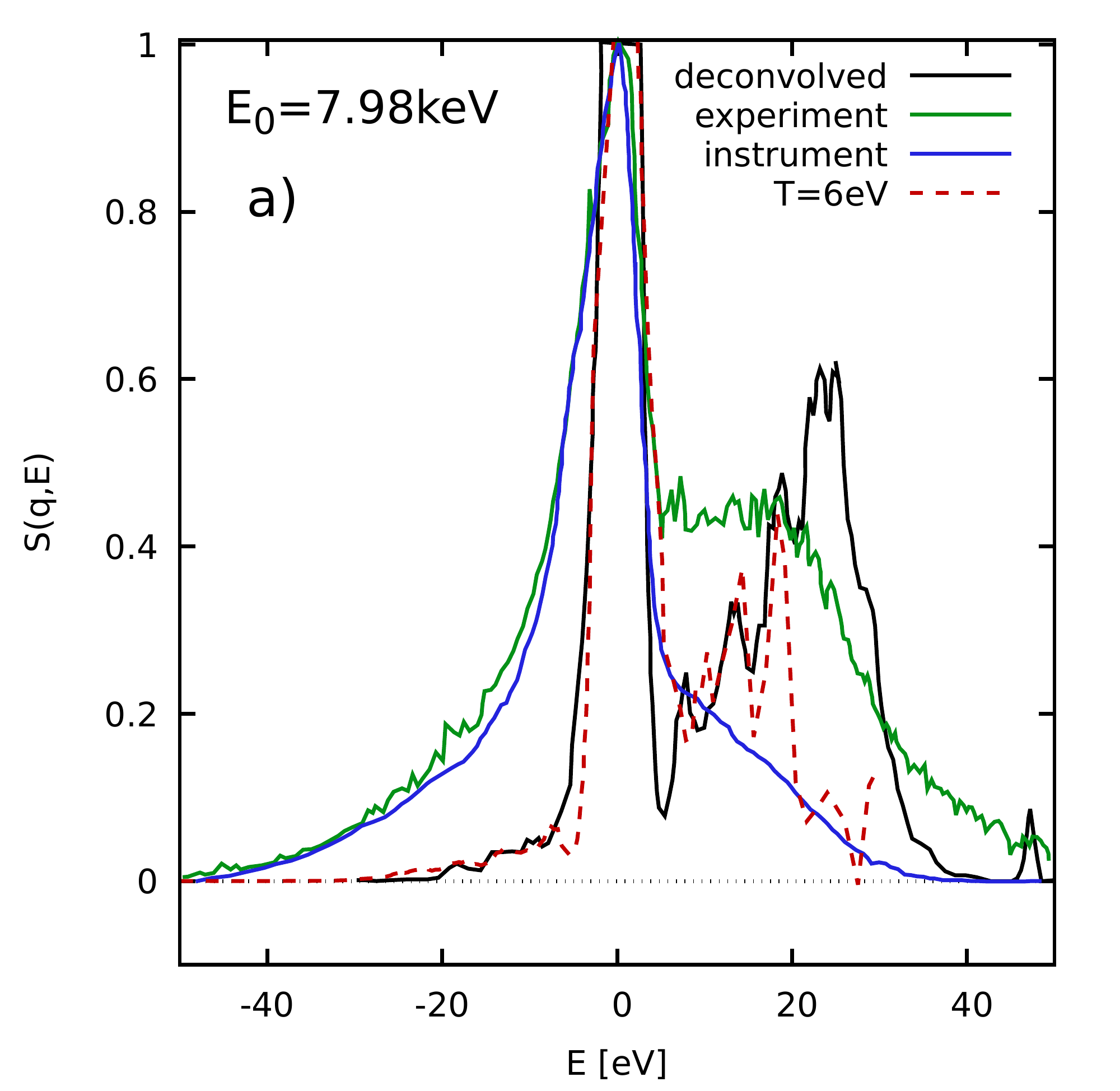}
\includegraphics[width=0.39\textwidth]{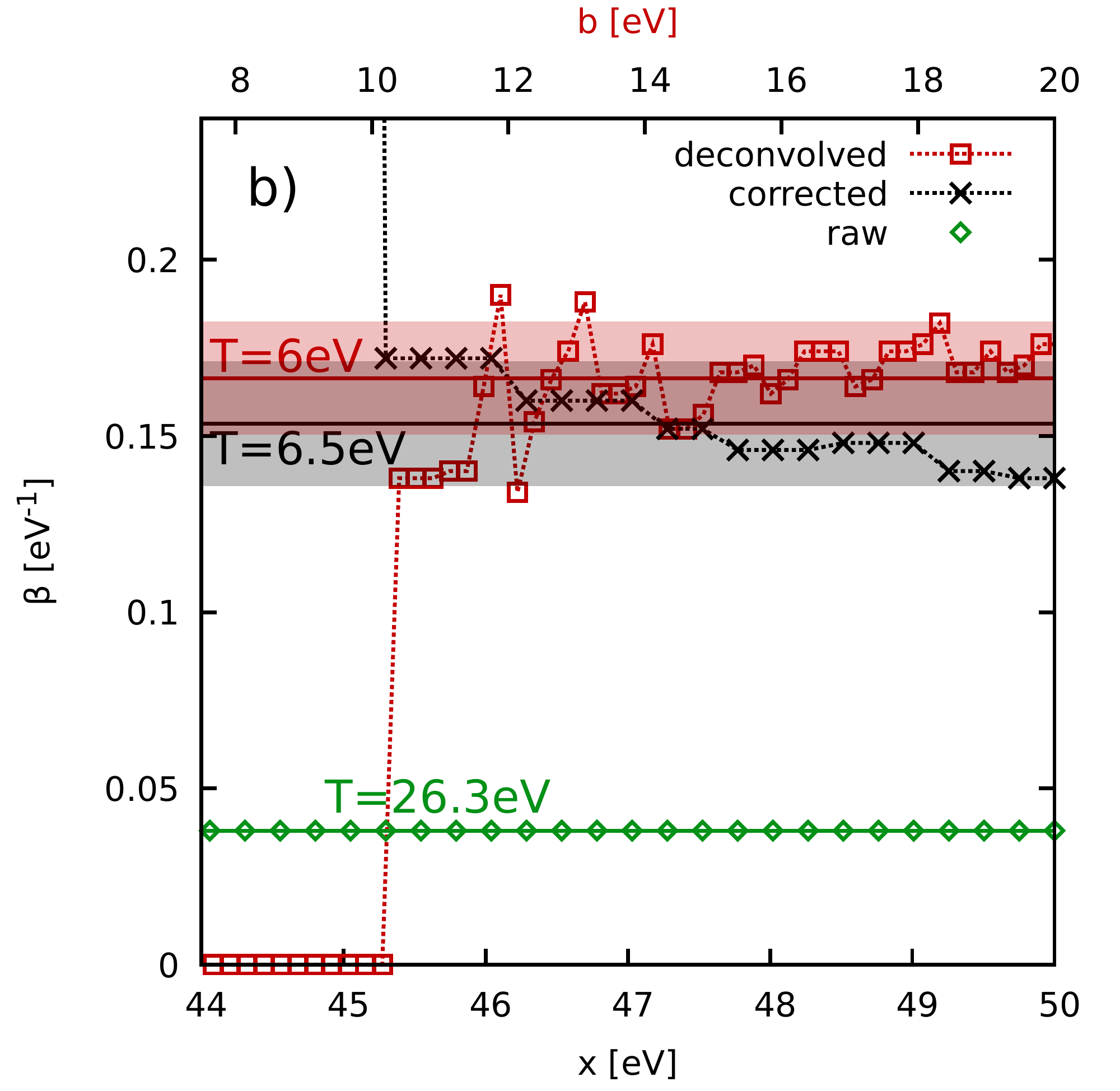}
\caption{\label{fig:Sperling}
Temperature diagnosis of warm dense aluminium. a) XRTS measurement by 
Sperling \emph{et al.}~\cite{Sperling_PRL_2015} (green), deconvolved dynamic structure factor (black), detailed balance estimation of $S(\mathbf{q},E)$ using the nominal value for the temperature of $T=6\,$eV (dashed red), and instrument function (blue); b) Convergence of our model-free temperature diagnostics with respect to the integration boundary $x$ of $\mathcal{L}\left[\dots\right]$ of the deconvolved data (red) [with the boundary $b$ being shown on the top abscissa], and Eq.~(\ref{eq:convolution}) with (black) and without (green) taking into account the effect of the instrument function $R(E)$.
}
\end{figure*}

\textbf{Aluminium experiment.}
As a second example, we consider the experiment with isochorically heated aluminium by Sperling \emph{et al.}~\cite{Sperling_PRL_2015} in Fig.~\ref{fig:Sperling}. This case has the considerable advantage that deconvolved data for $S(\mathbf{q},E)$ are available, see the black curve in panel a); the green and blue curves show the measured XRTS signal and instrument function, respectively. In the original publication, Sperling \emph{et al.}~\cite{Sperling_PRL_2015} have found a temperature of $T=6\,$eV based on a detailed balance evaluation of $S(\mathbf{q},E)$. Indeed, the corresponding red curve that has been obtained as $S_\textnormal{DB}(\mathbf{q},E) = S(\mathbf{q},-E)e^{-E/6\textnormal{eV}}$ is in excellent agreement to the deconvolved data in the range of $E\lesssim20\,$eV; the final peak around $E=30\,$eV
is likely absent from the negative energy range due to its vanishing amplitude in the deconvolved $S(\mathbf{q},E)$.
On the other hand, the original interpretation of the XRTS data has subsequently been disputed by independent groups on the basis of time-dependent density functional theory calculations and a model exchange--correlation kernel that has been constructed for the case of a uniform electron gas~\cite{Mo_PRL_2018,Dornheim_PRL_2020_ESA}. Specifically, these works have postulated substantially lower temperatures in the range of $T=0.3-2\,$eV, and hitherto no decisive conclusion had been reached.

In Fig.~\ref{fig:Sperling} b), we show the results of our new temperature diagnostic as a function of the integration range $x$. Specifically, the black crosses show our evaluation of Eq.~(\ref{eq:convolution}) taking into account the instrument function, and the green diamonds have been obtained without this correction. Evidently, the broadening of the XRTS signal by the instrument function plays a decisive role in this data, and leads to a five fold increase in the respective temperature. For the properly corrected data, we find a temperature estimate of $T=6.5\pm0.5\,$eV, which confirms the previous estimation by Sperling \emph{et al.}~\cite{Sperling_PRL_2015}. We also directly compute $F(\mathbf{q},\tau)$ from the deconvolved data for $S(\mathbf{q},E)$ via Eq.~(\ref{eq:ITCF}). The results are shown as the red squares, where the upper integration range is denoted as $b$ shown on the top abscissa. From panel a), it is clear that the integration only makes sense for $\lvert E \rvert\lesssim 20\,$eV, as no significant signal exists in the deconvolved data for $E<-20\,$eV.
This analysis gives us a temperature of $T=6\pm0.5\,$eV, and thereby further substantiates our calculation.
Therefore, our present analysis has shown that the temperature estimation in the original Ref.~\cite{Sperling_PRL_2015} is not an artefact due to the numerical deconvolution, as the direct interpretation of the XRTS signal via Eq.~(\ref{eq:convolution}) gives the same outcome.

\begin{figure*}\centering
\hspace*{-0.0cm}\includegraphics[width=0.39\textwidth]{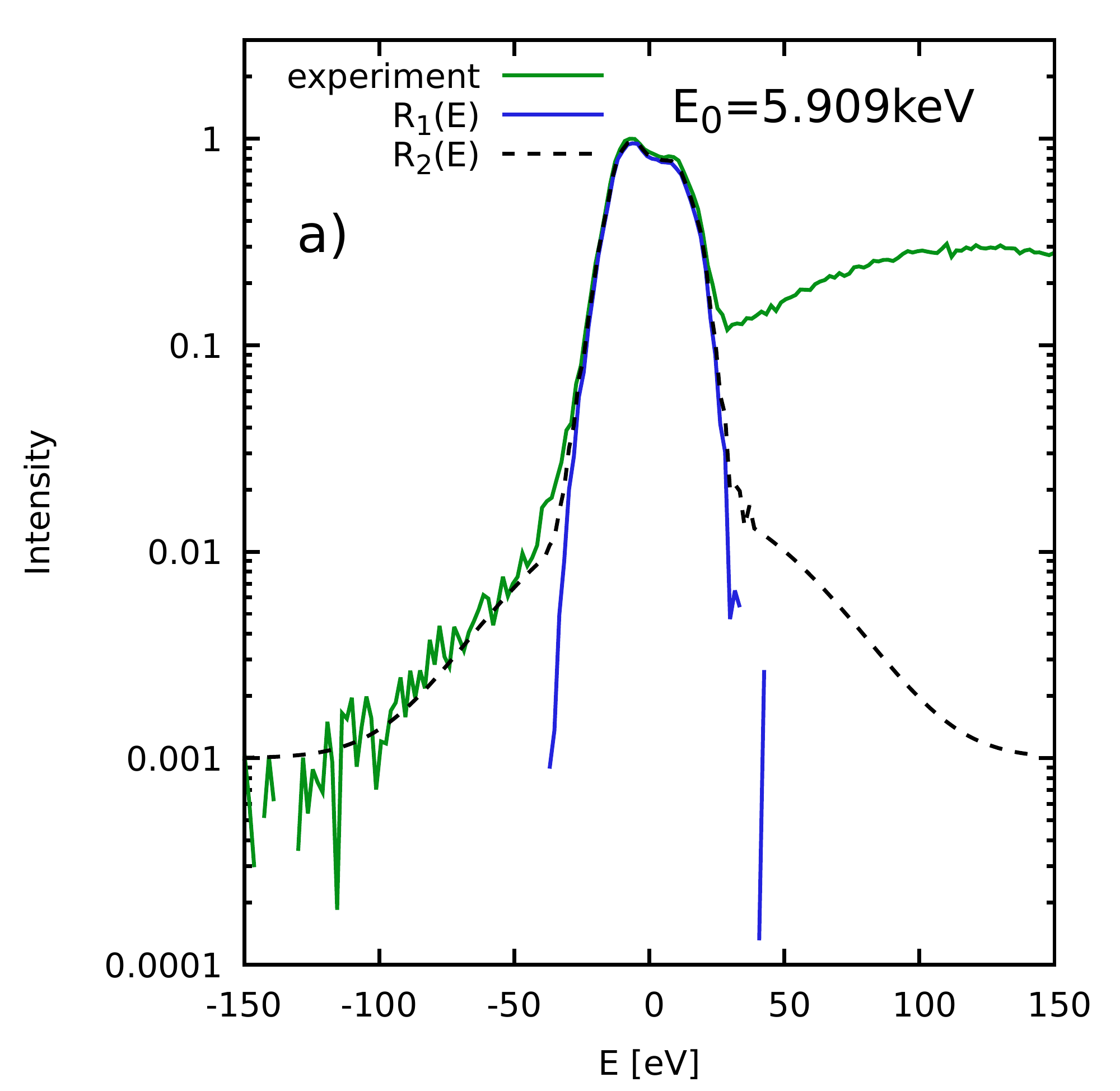}
\includegraphics[width=0.39\textwidth]{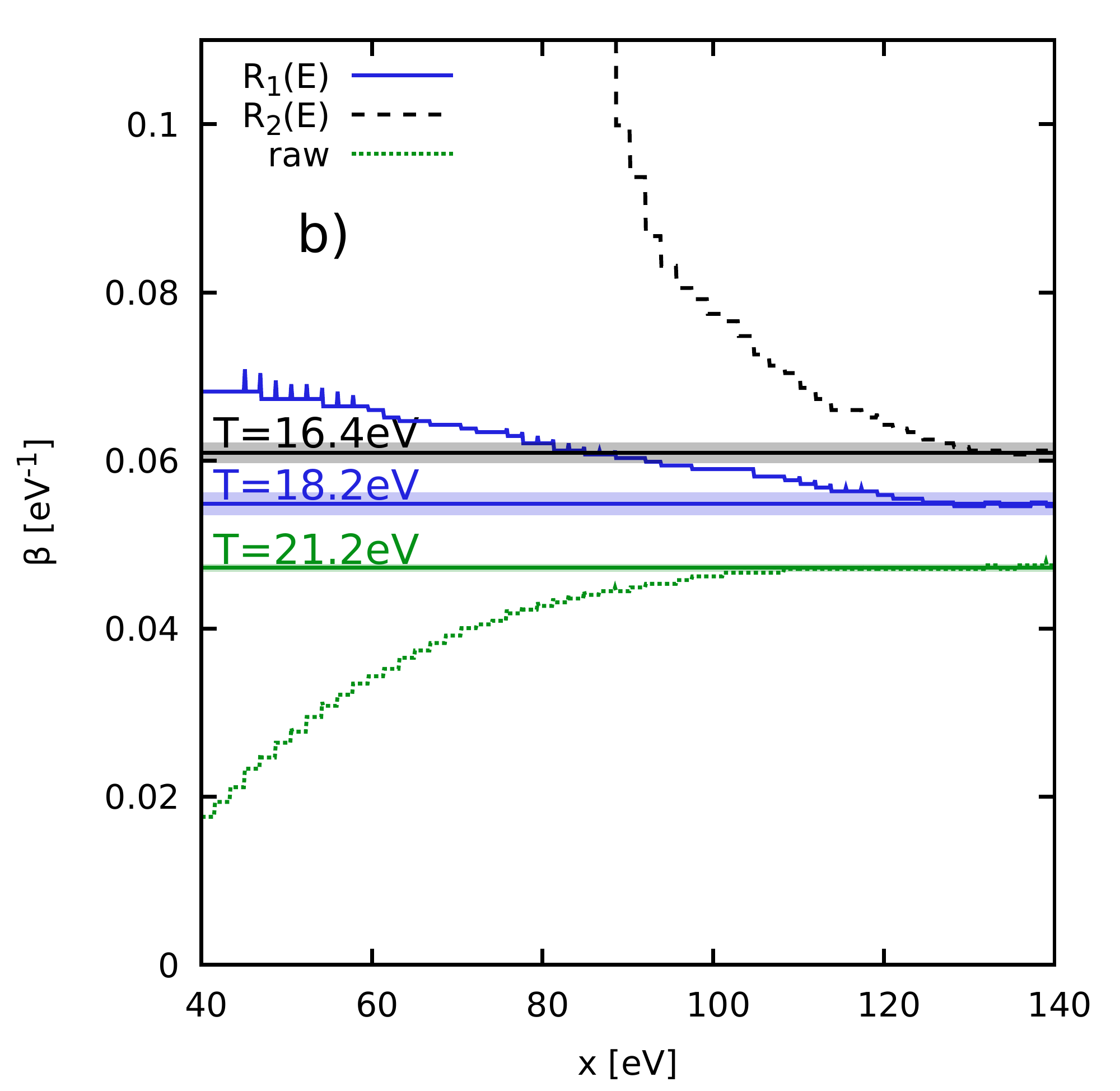}
\caption{\label{fig:Kraus}
Temperature diagnosis of warm dense graphite. a) XRTS measurement by Kraus \emph{et al.}~\cite{kraus_xrts} (green) and possible instrument functions $R_1(E)$ (blue) and $R_2(E)$ (dashed black) shown on a semi-logarithmic scale; b) Convergence of our model-free temperature diagnosis with respect to the integration boundary $x$ with (blue and black) and without (green) taking into account the instrument function.
}
\end{figure*} 

\textbf{Graphite experiment.}
As the final example, we re-examine the recent experiment on warm dense graphite by Kraus \emph{et al.}~\cite{kraus_xrts} in Fig.~\ref{fig:Kraus}. In this case, accurate data is available over three orders of magnitude in the measured XRTS signal (green). In addition, the solid blue and dashed black curves show two different models for the instrument function. In fact, this uncertainty regarding the true $R(E)$ has led to an uncertainty in the temperature of $\sim50\%$ based on the applied approximate Chihara models in Ref.~\cite{kraus_xrts}.

The outcome of our new temperature diagnostic is shown in Fig.~\ref{fig:Kraus} b). The green curve has been obtained without any correction due to the instrument function, leading to the biased temperature of $T=21.2\,$eV; this is very close to the value of $T=21.1\,$eV given by Kraus \emph{et al.}~\cite{kraus_xrts} based on the Chihara decomposition. Using the narrow instrument function $R_1(E)$ to compute the denominator in Eq.~(\ref{eq:convolution}) leads to the blue curve, with an estimated temperature of $T_1=18.2\,$eV. Using the broader $R_2(E)$ (truncated at $E\pm90\,$eV, as the constant asymptotes given in Ref.~\cite{kraus_xrts} are clearly unphysical and would lead to a divergent $\mathcal{L}\left[R_2(E)\right]$) for the correction gives the black curve, resulting in a second estimate of $T_2=16.4\,$eV. Evidently, the main source of uncertainty in the interpretation of this experiment is indeed given by the unclear shape of the instrument function. We therefore highlight the importance to accurately determine $R(E)$ in future experiments. At the same time, we note that our analysis suffers substantially less severely from this drawback compared to the original, Chihara model based interpretation, and we give our final estimate for the temperature as $T=18\pm2\,$eV.

\textbf{Discussion.} In this work, we have presented a new, highly accurate methodology for the temperature diagnosis of matter at extreme densities and temperatures based on scattering measurements. In particular, our new paradigm does not depend on any model. Therefore, it is free from approximations, and the negligible computation cost makes it highly suitable for the on-the-fly interpretation of XRTS experiments at modern facilities with a high repetition rate such as the European XFEL~\cite{Tschentscher_2017,Yabashi2017}. Moreover, it is very robust with respect to the noise of the measured intensity, and completely circumvents the crucial problem of the deconvolution with respect to the instrument function $R(E)$.
The presented practical application of our technique has given new insights into the behaviour of different materials in the WDM regime, and has substantially reduced previous uncertainties.

We are convinced that our new methodology will open up unprecedented possibilities for the experimental investigation of matter at extreme conditions in a range of different contexts. For example, our framework has clear ramifications for the impact of the instrument function on the interpretation of the XRTS signal 
and, in this way, will guide the development of future experimental set-ups. In addition, accurate knowledge of the temperature can be used to inform the extraction of other important system parameters such as the electron density $n_e$ or the charge state $Z$.

A key strength of our approach is given by the fact that it is completely model free and therefore can be straightforwardly applied to arbitrarily complex materials.
For example, critical challenges on the path towards achieving high energy gain in ICF implosion experiments are the mitigation of hydrodynamic instabilities and achieving high fuel compression~\cite{LANDEN2020100755}. This requires an improved understanding of radiation transport and hence material opacities along the implosion pathway to improve predictive capabilities of implosion simulations as key information such as the ionization state at high compression are highly controversial~\cite{PhysRevE.94.011202,PhysRevResearch.2.023260}. This highlights the importance for accurate and robust temperature measurements in complex ablator materials at warm dense matter conditions, which will be enabled by our new technique.

Similarly, our idea will boost the burgeoning field of laboratory astrophysics, which is concerned with the study of highly complicated material mixtures at the conditions encountered at planetary interiors~\cite{Georg2007}. Finally, we note that our method can directly be used in a number of situations beyond WDM. This includes lower densities as they occur in magnetic fusion reactors, where optical Thomson scattering is applied to determine the temperature~\cite{Froula_2006,Pasch_2016}.




\backmatter
\section*{Methods}

\bmhead{Symmetry of the imaginary-time intermediate scattering function}\label{sec:supplement}

The symmetry of the imaginary-time intermediate scattering function $F(\mathbf{q},\tau)$ directly follows by inserting the detailed balance relation of $S(\mathbf{q},E)$ into Eq.~(\ref{eq:ITCF}) from the main text,
\begin{eqnarray}\nonumber
F(\mathbf{q},\tau) &=& \int_{-\infty}^\infty \textnormal{d}E\ S(\mathbf{q},E) e^{-E\tau} \\\nonumber &=&
\int_0^\infty \textnormal{d}E\ S(\mathbf{q},E)\left\{ e^{-E\tau} + e^{-E(\beta-\tau)} \right\}\\
 &=& F(\mathbf{q},\beta-\tau)\nonumber\ .
\end{eqnarray}

\bmhead{Acknowledgments}

This work was partly funded by the Center for Advanced Systems Understanding (CASUS) which is financed by Germany's Federal Ministry of Education and Research (BMBF) and by the Saxon Ministry for Science, Culture and Tourism (SMWK) with tax funds on the basis of the budget approved by the Saxon State Parliament.  The work of T.~D\"oppner was performed under the auspices of the U.S. Department of Energy by Lawrence Livermore National Laboratory under Contract No. DE-AC52-07NA27344.


\bibliography{sn-bibliography}


\end{document}